\documentclass{article}
 
\usepackage{psfig} 
\usepackage{a4wide}  
\usepackage{verbatim} 

\newcommand {\be} {\begin{equation}}    
\newcommand {\bea} {\begin{eqnarray} \nonumber }    
\newcommand {\ee} {\end{equation}}    
\newcommand {\eea} {\end{eqnarray}}    
\newcommand {\fig}[1]{Fig.~\ref{#1}}

\newcommand {\nn} {\nonumber}     
\renewcommand {\d}{\partial}    
\newcommand{\prt}[1]{{(#1)}}

\def\tkappa{\tilde{\kappa}}    
\def\(({\left(}    
\def\)){\right)}    
\def\[[{\left[}    
\def\]]{\right]}    
\def \<{\langle}    
\def \>{\rangle}    
\def\N{{\cal N}}    
    
\begin{document}    
\bibliographystyle{prsty}    
\title{Replica Treatment of the Calogero--Sutherland Model }

    
\author{Dimitry M. Gangardt  and Alex Kamenev \\ \em Department of Physics, Technion, 32000 Haifa, Israel}    
\date{\today}

\maketitle    
    
\begin{abstract}   
Employing Forrester--Ha method of Jack polynomials, we derive an   
integral identity connecting certain  $N$--fold coordinate average   
of the Calogero--Sutherland model  with the $n$--fold replica   
integral.  
Subsequent analytical continuation   to non--integer $n$ leads to   
asymptotic expressions for the (static and dynamic) density--density   
correlation function of the model  as well as the  Green's function  
for an arbitrary coupling constant  $\lambda$.  
\end{abstract}

\section{Introduction}    
\label{s1}

The  model  of one--dimensional particles interacting     
via the inverse square potential, known as the Calogero--Sutherland model (CSM)   
\cite{CS}, has attracted a great deal of attention     
\cite{Forrester,Simons,Haldane,Ha}.   It has long been realized     
\cite{CS}    
that for the three special values of the interaction parameter     
$\lambda = 1/2,1,2$ the CSM is closely related to the     
random matrix theory (RMT) \cite{Mehta} of the      
three Wigner--Dyson  ensembles: orthogonal, unitary and     
symplectic respectively. Recently, using the method of Jack polynomials,    
Forrester \cite{Forrester} found the correlation functions  of the CSM for any integer   
$\lambda$.    
The method was latter generalized by Ha      
\cite{Ha}    
to any {\em rational} $\lambda=p/q$, where $p$ and $q$ are mutual primes.      
In the latter case the density--density correlation function of the model     
may be written as the $(p+q)$--fold integral over momenta of $p$ hole and    
$q$ particle excitations    
\cite{Haldane,Ha}.    
This seemingly makes the answer a  highly irregular function of the coupling     
strength $\lambda$.       
    
In this paper we employ the {\em replica} method, borrowed  from the  theory of disordered systems    
\cite{Edwards}     
and specifically RMT    
\cite{Zirnbauer,Kamenev,Lerner},    
to treat the CSM at an {\em arbitrary} value of $\lambda$.     
The idea is to write the particle density operator    
as  the $n$--th power of the characteristic polynomial     
in the limit $n\to 0$.     
In the context of the CSM, this appears to be a particularly simple     
operator which creates at most $n$ holes and no other excitations \cite{Ha}.    
As a result, for  an {\em integer} $n$, the correlation function of the      
characteristic polynomials     
may be expressed  as an     
$n$--fold integral over hole momenta for {\em any} coupling constant,    
$\lambda$.      
    
Mathematically, this statement boils down to the powerful     
asymptotic ($N\to \infty$)     
identity between the $N$--fold coordinate average of the      
$n$--th power of the characteristic polynomial       
and the $n$--fold momenta integral. For $\lambda=1$ (unitary RMT) such      
integral representation has been obtained earlier by  Verbaarschot and Zirnbauer    
\cite{Zirnbauer}    
by  mapping the problem onto the non--linear $\sigma$--model.   
We derive here the     
identity valid for any $\lambda$ employing the Forrester--Ha     
method of Jack     
polynomials. As of today we are not aware if this identity has a geometrical    
or field--theoretical derivation apart from $\lambda=1/2,1,2$.     
The simplicity of the final expression suggests the possibility that such a derivation     
exists.     
We stress, however, that for any $\lambda$ the identity has clear     
quantum--mechanical    
interpretation. We hope also that it may prove to be useful in the theory     
of the Coulomb gases.      
    
The  calculation requires evaluation  of the resulting $n$--fold     
integral for any integer $n$.     
We accomplish this goal in  the asymptotic regime $x\gg 1$     
($x$ is the scaled distance) by summing up contributions on $n+1$ saddle     
points     
($n-1$ of these saddle points may be dubbed as ``replica non--symmetric'').    
The subsequent analytic continuation, $n\to 0$, is    
analogous to  the one employed recently in the RMT context    
\cite{Kamenev,Lerner}. Some mathematical aspects of this procedure     
are not completely clear at the moment \cite{Zirnbauer02}.   
For rational coupling constants $\lambda$ our procedure is in    
exact agreement with the existing results  \cite{Ha,Zuk} and   
allows one  to obtain them in a much simpler way.   
In addition it leads to expressions formally valid for any $\lambda$.

The paper is organized as follows. In section \ref{s2} we define the     
CSM and introduce the  replica method. In section    
\ref{s3} we derive the integral identity discussed above. Section     
\ref{s4} is devoted to the calculation of the $n$--fold integral in the   
large $x$ limit and     
analytical continuation, $n\to 0$.    
Finally, our results and some     
conclusions are summarized in section \ref{s5}. Technical details are     
presented in three appendices.

\section{CSM and replica}    
\label{s2}    
    
\subsection{Eigenstates of CSM}

Consider  $N$ interacting particles on the unit circle.    
The CSM \cite{CS} is defined by the following Hamiltonian:    
\be    
H = \sum_{j=1}^{N}\left(z_j\d_j\right)^2+     
2 \sum_{j<i}\frac{\lambda(\lambda-1) }{|z_j-z_i|^2}\, ,    
                                                     \label{ham}    
\ee    
where $z_j=e^{i\theta_j}$ are the coordinates of the $j$-th     
particle,  $\d_j=\d/\d z_j$     
and $\lambda$ is the coupling constant.     
The first term is the usual kinetic energy, while the second one    
describes inverse square interactions. The spectrum and the many--body     
eigenfunctions of the model may be found exactly     
using the theory of the symmetric functions    
\cite{Jack,Macdonald}.     
For the sake of completeness and to set     
notations we present here the basic    
features of this solution (for more details see Appendix \ref{app1}).

\begin{figure}     
\centerline{a) \psfig{figure=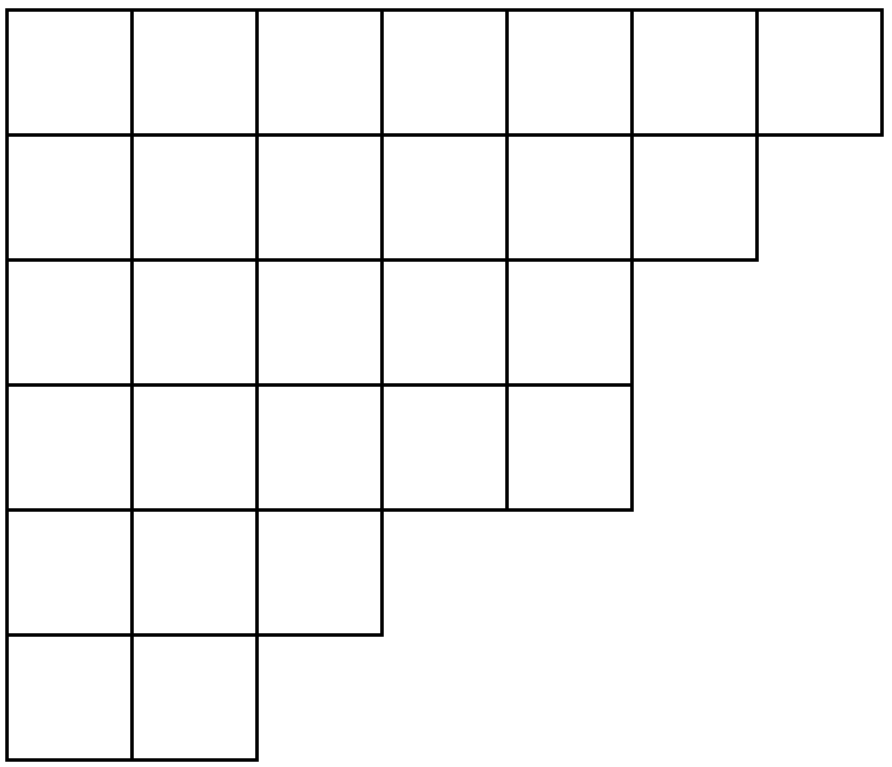,height=6cm} b) \psfig{figure=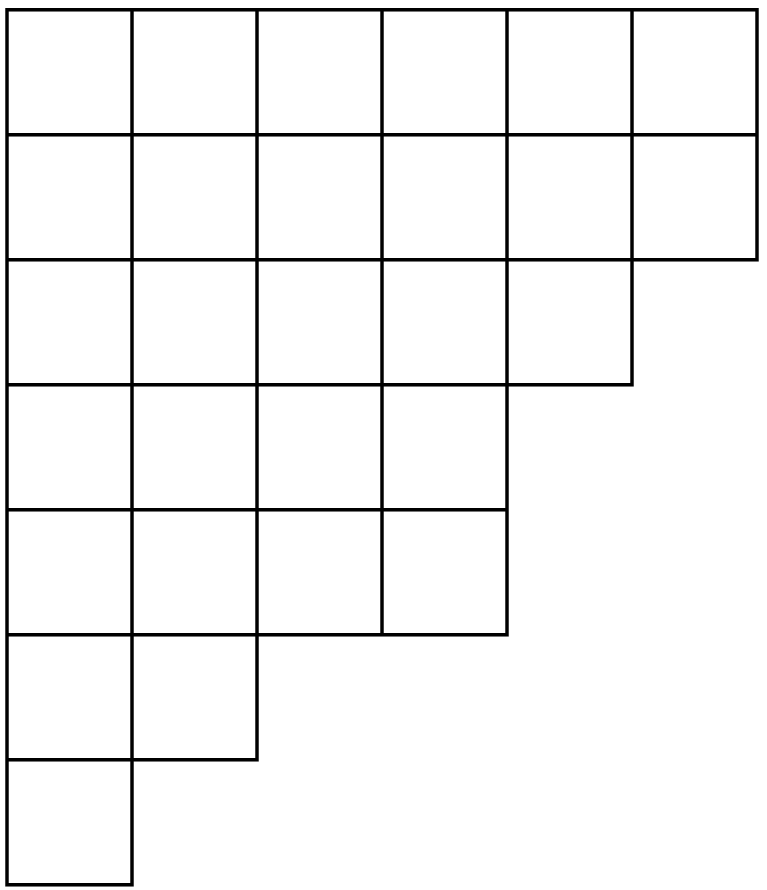,height=7cm} }    
\vspace{20pt}    
\caption{a) Young diagram for partition $\prt{\kappa}=\prt{7,6,5,5,3,2}$ representing an 
excitation of particles in the Calogero-Sutherland model. b) The Young diagram for the dual partition 
$\prt{\tkappa}=\prt{6,6,5,4,4,2,1}$ representing the excitation of holes.}     
\vspace{20pt}    
\label{yd}  
\end{figure}

The ground state wave-function of the CSM  \cite{CS} is \cite{footgs}    
\be    
\<z|0\>=\Psi_0 (z) = \sqrt{ \N}\,     
\left[ \Delta_N(z)\right]^\lambda   \, ,    
                                                        \label{gs}    
\ee    
where     
\be     
\Delta_N(z)  \equiv \prod_{j<i} \left(z_i-z_j\right)    
                                                \label{vandermonde}    
\ee    
is the $N\times N$ Vandermonde determinant and   the   
normalization constant is given by     
\be     
{\cal N } = \N(\lambda,N)= (2\pi)^{-2N}     
\frac{\Gamma^N (1+\lambda)}{\Gamma (1+\lambda N)}\, .    
                                              \label{normalization0}    
\ee    
Excited states  are     
labeled by the quantum numbers (essentially momenta)     
$\prt{\kappa} = (\kappa_1,\kappa_2,\ldots,\kappa_N)$, where     
the non--negative integers $\kappa_j$ are ordered as     
$\kappa_1 \geq \ldots\geq \kappa_N \geq 0$.     
In the mathematical literature the quantum numbers     
$\prt{\kappa}$   are called {\em partitions} \cite{Andrews} and    
represented by the {\em Young diagrams}, with $\kappa_j$  cells     
in $j$-th row. The example of the Young diagram for  the partition     
$\prt{\kappa}=\prt{7,6,5,5,3,2}$     
sometimes written as $\prt{7,6,5^2,3,2}$  is     
presented in \fig{yd}.     
In our case number of rows in a  Young diagram cannot exceed $N$,    
whereas number of columns is unlimited.  The  state $\prt{\kappa}$ has well defined total 
momentum, whose value is given by  the number of cells in the corresponding Young diagram     
and denoted as $|\kappa|\equiv \sum_{j=1}^N \kappa_j$.     
The coordinate representation wave-function of the  excited state corresponding to a given partition     
$\prt{\kappa}$ may be written as     
\be    
\<z|\prt{\kappa}\>=\Psi_{\prt{\kappa}} (z)     
= \sqrt{\N_\prt{\kappa}} J^{1/\lambda}_{\prt{\kappa}}(z)  \times     
\left[ \Delta_N(z) \right]^\lambda \; ,    
                                                             \label{es}    
\ee    
where    the symmetric     
functions   
$J^{1/\lambda}_\prt{\kappa} (z) =   
J^{1/\lambda}_\prt{\kappa} (z_1,\ldots,z_N)$   
are known as Jack polynomials \cite{Jack,Macdonald}.    
Definitions and some properties of Jack polynomials along with the     
normalization constants, $\N_\prt{\kappa}$,      
are presented in Appendices \ref{app1}, \ref{app2}.

\subsection{Correlation functions}

We focus first on  the ground state, static density--density correlation     
function of the  CSM:    
\be    
R(\theta)\equiv   
\rho_0^{-2} \<0|\tilde{\rho}(\theta)\,\tilde{\rho}(0)|0\>\, ,    
                                                         \label{ddcf}    
\ee    
where $\tilde{\rho}(\theta)=\rho(\theta)-\rho_0$ denotes the fluctuating     
part    
of the density operator $\rho(\theta)=\sum_j \delta (\theta-\theta_j)$    
and  $\rho_0 = N/2\pi$ is the mean density. Due to the     
translational invariance of the model     
this correlation function depends only on the relative angle     
between the two points.  One may  represent the     
fluctuating part of the density operator as the Fourier series:    
\be    
\tilde{\rho}(\theta) = \sum_{k\neq 0} \rho_k e^{-ik\theta} \, ,  
                                               \label{four}    
\ee    
where the Fourier coefficients of the density,     
$\rho_k=\bar\rho_{-k}$,     
are given by    
\be    
2\pi\rho_k  =  \sum_{j=1}^N e^{ik\theta_j} =     
 \sum_{j=1}^N z_j^k\, .    
                                               \label{fourcoeff}    
\ee    
In the theory of symmetric functions the Fourier coefficients   
$2\pi\rho_k$    
are referred to as {\em  power sums} of variables $z_j$.     
Since the Jack polynomials constitute a complete basis in the ring     
of symmetric polynomials the power sums $\rho_k$  can be decomposed     
into linear combinations of     
$J^{1/\lambda}_\prt{\kappa} (z)$, which is the basis of Ha's     
\cite{Ha}    
approach.  Here we employ a different method.    
    
Let us define the characteristic polynomial (or vertex) operator as     
\be    
Z(\theta) \equiv  
\prod_{j=1}^N \left(1-e^{-i(\theta-\theta_j)}\right) =     
\prod_{j=1}^N\left(1-e^{-i\theta} z_j\right)\, .    
                                                           \label{z}    
\ee    
In the RMT this object is known as the characteristic polynomial of the unitary 
matrix with eigenvalues $z_j$ or spectral determinant.      

Rewriting the characteristic polynomial in the form    
\be    
Z(\theta)=\exp\left\{ \sum_{j=1}^N \ln(1-e^{-i\theta} z_j)\right\} =    
\exp\left\{    
-2\pi \sum_{k=1}^\infty \frac{\rho_k e^{-ik\theta} }{k}     
\right\}\, ,    
                                                          \label{z1}    
\ee    
one notices that     
the fluctuating part of the density is given by    
\be    
\tilde{\rho}(\theta) = \frac{1}{\pi} \partial_\theta\mbox{Im}     
\ln Z(\theta) \; .    
                                                       \label{rhoz}    
\ee    
To ensure convergence of the series in Eq.~(\ref{z1}) one should assume     
an infinitesimal negative imaginary part of $\theta$:     
$\theta\to \theta-i0$.     
The correlation function Eq.~(\ref{ddcf}) may be expressed  therefore through     
the correlation  of  logarithms of the spectral determinants:    
\be    
R(\theta) =     
-\frac{\rho_0^{-2}}{2\pi^2}\, \partial_\theta^2\,    
\mbox{Re}\, \<0|\ln \bar{Z}(\theta)\ln Z(0)|0\> \; .    
                                                       \label{lnlncf}    
\ee    
The  replica trick is to write  the  logarithm as     
\be    
\ln Z = \lim_{n\rightarrow 0} \frac{1}{n}\left(Z^n -1\right)\; .    
                                                         \label{replica}    
\ee    
As a result one finally obtains for the correlation function     
\be    
R(\theta)=- \lim_{n\to 0} \frac{1}{2\pi^2n^2} \,   
{\partial^2 \over \partial(\rho_0\theta)^2}   \,     
\mbox{Re}\,  \<0|\bar{Z}^n (\theta) Z^n (0)|0\> \, ,  
                                                         \label{repddcf}    
\ee    
If one was able to calculate the expectation value on the r.h.s.   
for any real     
$n$ the limiting procedure would be straightforward. The non--trivial     
aspects of the replica method show up if the calculation is possible     
only for positive   
\cite{footbosons}   
integer $n$. As we shall see in the next section this    
is indeed the case.

\section{Correlations of characteristic polynomials}    
\label{s3}    
    
In this section we calculate the large $N$ limit of the correlation     
function of the     
characteristic polynomials for integer $n$.      
In the coordinate representation it takes the form of the     
following $N$--fold integral    
\be      
\<0|\bar{Z}^n (\theta) Z^n (0)|0\> = \N      
\int\limits_0^{2\pi}\!  d^N \theta \,     
|\Delta_N(z)|^{2\lambda}     
\left[ \prod_{j=1}^N \left(1-e^{i\theta}\bar{z}_j\right)    
 \left(1-z_j\right)\right]^n \, ,     
                                                    \label{corrint}    
\ee    
where the normalization constant is given by Eq.~(\ref{normalization0}).     
    
Consider the ground state expectation value of the $s$-th power of the   
characteristic polynomials, where $s$ is an arbitrary real number.  
Inserting resolution of unity,  one finds     
\be    
\<0|\bar{Z}^s (\theta) Z^s(0)|0\> =  \sum_\prt{\kappa}    
\<0|\bar{Z}^s (\theta)|\prt{\kappa}\>      
\<\prt{\kappa}| Z^s(0) |0\>     
=  \sum_\prt{\kappa}        
|\<\prt{\kappa}| Z^s(0) |0\>|^2 e^{i|\kappa|\theta}    \, .    
                                                      \label{corr0}    
\ee    
where the sum runs over {\em all} possible partitions (with number of     
rows less than $N$). The last equality in Eq.~(\ref{corr0}) is a trivial   
consequence of the translational invariance and the fact that the state   
$|\prt{\kappa}\>$ carries 
momentum $|\kappa|$.   
Since $Z^s(0)$ is a  symmetric function of $z$ it may be expanded   
in the basis of Jack polynomials. Employing     
the orthogonality relation for Jack polynomials, one obtains the matrix   
elements    $\<\prt{\kappa}| Z^s(0) |0\>$.   
Details of this calculation are given in Appendix \ref{app2}.   
The result is  
\be    
\<0|\bar{Z}^s (\theta) Z^s(0)|0\> =  \sum_\prt{\kappa}     
|A_\prt{\kappa}(\lambda,N)|^2\,   
\((\{-s\}^\lambda_\prt{\kappa}\))^2    
e^{i|\kappa|\theta}\, ,    
                                                          \label{corr}    
\ee    
where $A_\prt{\kappa}(\lambda,N)$ is an $s$--independent   
factor, which is given in Appendix \ref{app2} and   
the symbol $\{-s\}^\lambda_\prt{\kappa}$ is defined as    
\be    
\{-s\}^\lambda_\prt{\kappa} = \prod_{j,a\in \prt{\kappa}}    
\Big(-s-\lambda (j-1)+(a-1)\Big)\, .     
                                                     \label{dcurly}    
\ee    
Here the product runs over all cells of the Young diagram for a given     
partition $\prt{\kappa}$ with     
$j=1,\ldots,N$ being the vertical coordinate and  $a=1,2,\ldots$ the horizontal    
coordinate of the cell.

At this point the advantage of the replica approach may be appreciated.    
One notices that for $s=n$ -- positive integer,  the factor     
$\{-n\}^\lambda_\prt{\kappa}$ defined by Eq.~(\ref{dcurly}) is zero     
for any partition $\prt{\kappa}$ which has more than $n$ columns in its     
Young diagram. Indeed, for $j=1$ (the upper row) and $a=n+1$ the factor   
on the r.h.s of     
Eq.~(\ref{dcurly}) nullifies for {\em any} $\lambda$. As a result the set     
of partitions which contribute to the correlation function,     
Eq.~(\ref{corr}), is greatly reduced. It contains only partitions which     
have not more than $N$ rows and not more than $n$ columns in their Young    
diagrams.     
    
The similar drastic reduction in the number of contributing     
partitions takes place in the Ha's approach    
\cite{Ha}.    
However, in that case it happens only for rational $\lambda=p/q$ and     
the geometry of the relevant Young diagrams is determined by $p$ and $q$.     
In the replica approach the fact that $n$ is integer takes care of     
diagrams selection, whereas   $\lambda$ may be arbitrary. In other words,    
the operator $Z^n$ is a particularly simple one: by acting on the     
ground state it creates not more than $n$ hole--type excitations. As a result,    
number of partitions (excited states) needed to represent      
$Z^n|0\>$ is finite and each partition can be parametrized by the heights of its $n$ columns.      
    
The next step is to take the thermodynamic limit, $N\to \infty$. To this end     
we parameterize the relevant partitions by the height of the columns in their     
Young diagrams. This introduces the  {\em dual} partition     
$\prt{\tkappa} = \prt{\tkappa_1,\tkappa_2,\ldots,\tkappa_n}$ whose     
Young diagram is just a transposition of the diagram for the      
partition $\prt{\kappa}$. Since     
Young diagram of $\prt{\kappa}$ has no more than $N$ rows and no     
more than $n$ columns, 
$0\le \tkappa_a \le N$ and $a=1,2\ldots n$.    
Physically $\tkappa_a$ describe momenta of $n$ holes; each  cannot     
exceed $N$ since the ``Fermi'' sea is bound  (recall that $\kappa_j$     
describe  unbounded momenta of $N$ particles).       
In the limit $N\to \infty$ one may  pass from summations over  $\tkappa_a$    
in Eq.~(\ref{corr}) to integration over $k_a \equiv \tkappa_a/N$,     
where $0\leq k_a\leq 1$, simultaneously taking the limit in     
$A_\prt{\kappa}(\lambda,N\to\infty)$. The corresponding procedure is     
described in details in Appendix \ref{app3}. As a result one finds in     
the large $N$ limit:    
\be    
\<0|\bar{Z}^n (\theta) Z^n(0)|0\> =     
N^{{n^2\over\lambda}}    
{\cal C}(\lambda, n) \int\limits_0^1 \!    
d^n k      
\left|\Delta_n(k)\right|^{2/\lambda}     
\left[ \prod_{a=1}^n k_a (1-k_a) \right]^{1/\lambda-1}\!\!     
\exp\left \{iN\theta \sum\limits_{a=1}^n k_a \right\}\, ,     
                                                        \label{corrtherm}    
\ee    
where the  normalization constant is     
\be    
{\cal C}(\lambda, n) =      
\prod_{a=1}^n \frac{\Gamma(1+1/\lambda)}  
{\Gamma(1+a/\lambda)\Gamma(a/\lambda)}  \, .  
                                                          \label{c}    
\ee

Finally, employing Eqs.~(\ref{corrint}) and (\ref{normalization0}), one may     
express the result as the identity of the two integrals       
\begin{eqnarray}    
\lim_{N\to \infty}    
N^{-{n^2\over\lambda}}     
\N(\lambda,N)    
&&\int\limits_0^{2\pi}\!  d^N \theta\,     
|\Delta_N(z)|^{2\lambda}     
\left[ \prod_{j=1}^N \left(1-e^{2\pi i x/N}\bar{z}_j\right)    
 \left(1-z_j\right)\right]^n                      
                                               \nonumber \\    
=     
{\cal C}(\lambda,n)     
&&\int\limits_0^1 \!     
d^n k      
\left|\Delta_n(k)\right|^{2/\lambda}     
\left[ \prod_{a=1}^n k_a (1-k_a) \right]^{1/\lambda-1} \!\!    
\exp\left\{ 2\pi i x  \sum\limits_{a=1}^n k_a \right\}\, ,   
                                                \label{intid}    
\end{eqnarray}     
where we have introduced  the scaling variable   
$x \equiv \rho_0\theta =N\theta/(2\pi)$   
which measures  distance in units of the mean inter-particle separation,   
to emphasize that the left hand side has a well  defined      
thermodynamic limit.      
    
For $\lambda=1$ (unitary RMT), the  identity Eq.~(\ref{intid}) was derived by     
Verbaarschot and Zirnbauer   
\cite{Zirnbauer},   
using {\em fermionic} replica   
\cite{footbosons}  
and  mapping the random matrix problem onto the     
non--linear $\sigma$--model on the  coset space     
$U(2n)/U(n)\times U(n)$. In that formulation parameters $k_a$ play the     
role of the  (compact) angular coordinates on the coset. In a similar way    
one may  derive Eq.~(\ref{intid}) for $\lambda=1/2$ and $2$ (orthogonal and     
symplectic RMT) \cite{Zirnbauer01}. It would be extremely interesting     
to see if there is geometrical field--theoretical interpretation of     
Eq.~(\ref{intid}) for other $\lambda$.

The integral identity Eq.~(\ref{intid}) is exact for any $\theta$.    
In particular, for $x=0$ its right hand side    
may be evaluated  with the help of the  Selberg integral \cite{Mehta}     
leading to     
\be    
\lim_{N\to \infty}    
N^{-\frac{n^2}{\lambda}}    
\<0|\bar{Z}^n(0) Z^n (0)|0\> =    
\prod_{a=1}^n \frac{\Gamma(a/\lambda)}{\Gamma\Big((a+n)/\lambda\Big)}\, :    
                                                       \label{intid0}    
\ee    
result obtained previously by Keating \cite{Keating} for $\lambda=1$.    
For $x\neq 0$ the integral on the r.h.s. can not be evaluated     
exactly. In the next section we shall evaluate it in the asymptotic limit     
$x\gg 1$ employing the stationary phase  method as suggested by Forrester  \cite{Forrester}    
and Yurkevich and Lerner \cite{Lerner}.

\section{Replica Limit}  
\label{s4}    
  
For $x\gg 1$ the dominant contribution to the integral on the r.h.s. of   
Eq.~(\ref{intid}) comes from the end points of the interval.  We expand   
the integrand in $l$ variables about 1 and in remaining $n-l$ variables   
about 0 to leading order. Extending limits of integration to infinity,   
one obtains for the integral:   
\be  
\sum\limits_{l=0}^n {n\choose l}   
I_{n-l}(-2\pi i x) I_l(2\pi i x)   
e^{2\pi i x l} \, ,  
                                                  \label{split}  
\ee  
where  $I_m$ are     Selberg integrals \cite{Mehta} :  
\be  
I_m(z) = \int\limits_0^\infty\! d^m k   
|\Delta_m(k)|^{2/\lambda} \prod_{a=1}^m k_a^{1/\lambda-1}   
e^{-z k_a}  =   
z^{-{m^2\over \lambda} }   
\prod_{a=1}^m   
\frac{\Gamma(1+a/\lambda)\Gamma(a/\lambda)}{\Gamma(1+1/\lambda)} \, .  
					\label{il}  
\ee  
As a result, one obtains for the correlation function of the   
characteristic polynomials in the asymptotic limit   
$N\to \infty, x\gg 1$  
\be   
\<0|\bar{Z}^n (\theta) Z^n(0)|0\> = N^{n^2\over\lambda}  
\sum_{l=0}^{\infty} \left[F^l_n (\lambda)\right]^2  
\frac{e^{2\pi i x l} }  
{(-2\pi i x)^{(n-l)^2\over\lambda} (2\pi i x)^{l^2\over\lambda}   }\, ,  
  					\label{intsum1}  
\ee  
where   
\be  
F^l_n (\lambda) \equiv {n\choose l}  
\frac{\prod\limits_{a=1}^l\Gamma(1+a/\lambda)  
\prod\limits_{a=1}^{n-l}\Gamma(1+a/\lambda)}  
{\prod\limits_{j=a}^n\Gamma(1+a/\lambda)} =  
 {n\choose l}   
\prod\limits_{a=1}^l\frac{\Gamma(1+a/\lambda)}  
{\Gamma\Big(1+(n-a+1)/\lambda\Big)} \,   
					\label{fnl}  
\ee  
and $F_n^0(\lambda)=1$. We have extended summation over $l$ to infinity   
in Eq.~(\ref{intsum1}),   since for an integer $n$ the binomial coefficient   
${n\choose l}$ is identically zero for $l>n$.  
For RMT  ( $\lambda=1/2,1,2$ ) Eq.~(\ref{intsum1}) was derived recently   
from the saddle   
point treatment of the  non--linear $\sigma$--model   
\cite{Kamenev,Lerner}.   
In that case  the factor $F_n^l(\lambda)$ is     
the volume of the degenerate coset space manifold    
$G(n)/G(l)G(n-l)$, where $G$ is     
an appropriate symmetry  group, $G=Sp, U, O$ for $\lambda=1/2.1.2$ correspondingly.   
  
Up to this moment all our calculations were performed for   
integer $n$ only. Next we shall continue analytically  Eq.~(\ref{intsum1})   
to arbitrary real $n$ and then take the replica   
limit $n\to 0$. To this end we notice that   
\be    
{n\choose l}  
\equiv {\Gamma(1+n) \over \Gamma(1+l) \Gamma(1+n-l)}     
\to n{(-1)^{l+1} \over l}\, ,     \hskip 1cm n\to 0\,      
                                                     \label{Cnl1}  
\ee      
This expression  guarantees cancellation of the factor     
$n^{-2}$ in Eq.~(\ref{repddcf}) for small $n$. As a result one obtains   
for $l\geq 1$ and $n\to 0$   
\be     
F_{n}^{l}(\lambda) = n{(-1)^{l+1} \over l}    
\prod\limits_{a=1}^l     
{ \Gamma(1+a/\lambda)  \over \Gamma(1-(a-1)/\lambda)} +     
O(n^2)\, .    
                                                               \label{Fnl1}    
\ee    
Finally, employing  Eq.~(\ref{repddcf}) and taking the replica limit, one     
finds for the correlation function    
\be    
R(x) =  - {1 \over 2\pi^2 \lambda x^2} +    
2\sum\limits_{l=1}^\infty     
{d^2_l(\lambda) \over (2 \pi x)^{2l^2/\lambda} } \,     
\cos(2 \pi l x)  \, ,    
                                                 \label{R1}    
\ee    
where     
\be     
d_l(\lambda) =      
\frac{ \prod\limits_{a=1}^l \Gamma(1+a/\lambda)}    
{\prod\limits_{a=1}^{l-1} \Gamma(1-a/\lambda)} =   
\Gamma\((1+ {l\over\lambda}\))  
\prod\limits_{a=1}^{l-1}\(({a\over \pi \lambda}\))  
\sin\(({\pi a \over \lambda}\))\, \Gamma^2\(({a\over\lambda}\))   
\, .    
                                                  \label{al}    
\ee    
The first term on the r.h.s. of Eq.~(\ref{R1}) comes from the $l=0$   
contribution. It may be called replica symmetric, since all $n$ integrals   
are calculated around $k_a=0$ (the left  ``Fermi'' point). The oscillatory   
contributions on the r.h.s. of Eq.~(\ref{R1}) may be dubbed ``replica   
non--symmetric'', since in $l$ out of $n$ replica contribution   
to the integral comes from the vicinity of $k_a=1$ (the right``Fermi'' point).

For rational coupling constant $\lambda=p/q$   
the sine  function in Eq.~(\ref{al}) vanishes  for $a=p$. As a result,  
$d_l(p/q)=0$ for $l>p$ and therefore the series in   
Eq.~(\ref{R1}) contain only $p$ oscillatory components.  
For this case our results Eqs.~(\ref{R1}), (\ref{al}) coincide exactly with the  
asymptotic expression for the density--density correlation function obtained by  
Ha \cite{Ha}. In Ref.~\cite{Ha} the coefficients $d_l^2(\lambda)$ are given in   
terms of the  Dotsenko--Fateev integrals. One must employ the integral identity   
of Forrester and Zuk \cite{Zuk} to show the equivalence of coefficients in   
Ref.~\cite{Ha} and Eq.~(\ref{al}). In fact our approach together with Ref.~\cite{Ha} 
may be considered as an independent evaluation of some of the   
Dotsenko--Fateev integrals.   
We believe that for irrational $\lambda$ the series in Eq.~(\ref{R1}) is an   
asymptotic one, but its exact meaning and possible re-summation   
procedure requires more study.

\section{Discussion of the Results}    
\label{s5}

Let us briefly discuss the  generalization of the above procedure to the dynamical   
correlation function   
$R(x,t) = \rho_0^{-2} \<0|\tilde{\rho}(x,t)\,\tilde{\rho}(0,0)|0\>$.   
The characteristic polynomials and replica are introduced in exactly the   
same way  as for the static correlation function. One should take into   
account that operators at different times do not commute and that   
$\rho_{-k}(t)=\rho_k^\dagger(t)$. As a result one finds    
(cf. Eqs.~(\ref{repddcf}) and (\ref{corr0}) )   
\be   
R(x,t)=- \lim_{n\to 0} \frac{1}{2\pi^2n^2} \,   
{\partial^2 \over \partial x^2}   \,     
\sum_\prt{\kappa}        
|\<\prt{\kappa}| Z^n(0) |0\>|^2 \cos(|\kappa| x/\rho_0)   
e^{-it(E_\prt{\kappa} - E_0)}    \, ,     
                                                      \label{corrdyn}    
\ee    
where the energies of excited states, $E_\prt{\kappa}$, are given   
in Appendix \ref{app1}. The   
matrix elements of $Z^n(0)$ and their thermodynamic limit are exactly the   
same as for the  static case. The thermodynamic limit of the excitation   
energy in terms of hole momenta in the proper frame of reference is given  
by Eq.~(\ref{etk}). As a result, the integral on the r.h.s. of   
Eq.~(\ref{corrtherm}) acquires the factor   
$\exp\{-2\pi i[ v_s t \sum_a k_a(1-k_a)\mp x\sum_a k_a] \}$, where   
$v_s\equiv 2\pi \rho_0^2 \lambda$ is the sound velocity  
in the rescaled coordinates.   
  
Asymptotic calculation of the $n$--fold momenta integral in the limit   
$x\gg 1$ and/or $v_s t \gg 1$ follows closely the static case. The only   
difference is that the product of Selberg integrals   
$2I_{n-l}(-2\pi i x) I_l(2\pi i x)$ in Eq.~(\ref{split}) is substituted by   
$I_{n-l}(2\pi i [v_s t- x]) I_l(2\pi i [v_s t +x]) +(x\to -x)$.   
Performing the analytical continuation $n\to 0$ and employing   
Eq.~(\ref{corrdyn}), one finally finds for the dynamic correlation function   
\be    
R(x,t) =  - {1 \over 4 \pi^2 \lambda }  
\(( {1\over (x-v_s t)^2}  +  {1\over (x+v_s t)^2} \)) +  
2\sum\limits_{l=1}^\infty     
{d^2_l(\lambda) \over (4 \pi^2 [x^2 - v_s^2 t^2])^{l^2/\lambda} } \,     
\cos(2 \pi l x)  \, ,    
                                                 \label{R2}    
\ee    
where the coefficients $d_l(\lambda)$ are given by Eq.~(\ref{al}).     
  
The method also enables one to calculate the single particle Green function of the model. 
As it was shown in Ref.~\cite{Ha}  
the annihilation operator of particle  acting on the ground state of $N+1$ particles
is proportional to the action of 
$Z^{\lambda}$ on the $N$ particle ground state, {\it i.e.}
\be
\<0|\Psi^\dagger (x,t) \Psi (0,0)|0\> = (N+1) 
\frac{{\cal N}(\lambda,N+1)}{{\cal N}(\lambda,N)} 
\sum_\prt{\kappa}        
|\<\prt{\kappa}| Z^\lambda(0) |0\>|^2   
e^{i ( 2\pi |\kappa|/N-\pi\lambda ) 
x-i(E_\prt{\kappa} - \mu)t}    \, ,     
                                                      \label{green}    
\ee    
Therefore to evaluate the Green function  
one has to perform  analytical  continuation of (\ref{intsum1})   from integer $n$  
to $n=\lambda$. This yields in the large $N$ limit
\be
\<0|\Psi^\dagger (x,t) \Psi (0,0)|0\>= \frac{N}{2\pi}\sum_{l=0}^\infty 
\left[F^l_\lambda (\lambda)\right]^2 
\left(\frac{1}{2\pi i(v_s t-x)}\right)^\frac{(l-\lambda)^2}{\lambda} 
\left(\frac{1}{2\pi i(v_s t+x)}\right)^\frac{l^2}{\lambda} 
e^{i2\pi  (l-\lambda/2)x +i \mu t}
\label{green1}
\ee
where the coefficients $F^l_\lambda (\lambda)$ are still given by (\ref{fnl}) with $n\to \lambda$.
For rational $\lambda=p/q$ they vanish for $l>p$ as in the case of the density-density correlations,
so the sum in (\ref{green1}) contains a finite number of terms.

In summary, the correlation functions of the integer powers $n$ of the spectral determinants may be evaluated 
as a sum over partitions with no more than $n$ columns.  
In the thermodynamic limit such a sum takes a form of the $n$--fold integral, which may  
be evaluated for the large inter-particle separation. The subsequent analytical  
continuation to non--integer $n$ results in the density--density correlation  
function ($n\to0$) or in the single particle Green function ($n\to\lambda$). 
These calculations do not require rational values of the coupling constant  
$\lambda$ and are formally valid   for any $\lambda$. The final expressions, 
however, take the form of divergent series for irrational $\lambda$, whose exact  
meaning is  not clear at the moment. For rational $\lambda$ the results coincide  
with those of Ref.~\cite{Ha}.

Discussions with B. Altshuler initiated this project. We are also   
indebted to Y. Avron, J. Feinberg, S. Fishman, B. Simons M. Zirnbauer and A. Ludwig 
for valuable conversations.   
This research was supported in  part by  the U.S.--Israel Binational Science Foundation (BSF), 
by the Minerva   
Center for Non-linear Physics of Complex Systems, by the Israel  Science Foundation, by the Niedersachsen  
Ministry of Science  (Germany) and by the Fund for Promotion of Research at the   
Technion.

\appendix

\section{Eigenstates of the Calogero-Sutherland model}    
\label{app1}

Consider    action of the kinetic term of the Hamiltonian   
Eq.~(\ref{ham}) on the ground state, Eq.~(\ref{gs}), which is convenient   
to rewrite as $\exp\{\lambda\ln \Delta(z)\}$ :  
\be    
\sum_{j=1}^N \left(z_j\d_j\right)^2 [\Delta(z)]^\lambda =    
\sum_{j=1}^N\left[ \lambda \left(z_j\d_j\right)^2 \ln\Delta (z)+    
\lambda^2 \Big(z_j\d_j\ln\Delta (z)\Big)^2\right][\Delta(z)]^\lambda\, .    
						\label{kinvdm}    
\ee    
The logarithmic derivatives are evaluated as follows:  
\bea  
z_j\d_j \ln \Delta (z) &=&\sum_{i\neq j}\frac{ z_j}{z_j-z_i}\, ;  \\  
\left(z_j\d_j\right)^2 \ln\Delta (z)&=&  \sum_{i\neq j}\frac{ z_j}{z_j-z_i}-  
 \sum_{i\neq j}\frac{ z_j^2}{(z_j-z_i)^2} .  
						\label{logder}  
\eea  
As a result the action of kinetic term Eq.~(\ref{kinvdm}) is equivalent   
to the multiplication by the following factor    
\be  
\lambda(\lambda-1)\sum_{i\neq j} \frac{z_iz_j}{(z_i-z_j)^2}  
+\lambda^2 \left(\sum_{i\neq j\neq k}\frac{z_i^2}{(z_i-z_j)(z_i-z_k)}   
+\frac{N(N-1)}{2}\right)\, .  
						\label{kinvdm1}  
\ee  
The first term in this expression cancels exactly  the action of   
the interaction term of the Hamiltonian Eq.~(\ref{ham}), while the   
second term may be easily shown to be a constant.   
That proves that the wave-function Eq.~(\ref{gs}) is indeed an eigenstate of   
the Hamiltonian Eq.~(\ref{ham}) with the energy  
\be  
E_0 (\lambda) = \lambda^2 \left(\frac{N(N-1)(N-2)}{3}+  \frac{N(N-1)}{2}\right) =   
\frac{\lambda^2 N (N-1)(2N-1)}{6}\, ,  
						\label{e0}  
\ee   
which may  be considered as a sum of $N$ independent one-particle   
energies $\epsilon_j = \lambda^2 (N-j)^2$ for $j$ running from $1$ to  
$N$.   
The normalization of the ground state wave function is obtained   
straightforwardly with the help of the Selberg integral \cite{Mehta}  
\be  
\int\limits_0^{2\pi}\!   
\frac{d^N\theta}{(2\pi)^N}   
\prod_{j<i} \left|e^{i\theta_i}-e^{i\theta_j}\right|^{2\lambda} =   
 \frac{\Gamma (1+\lambda N)}{\Gamma^N (1+\lambda)}    
 						\label{norm}  
\ee  
resulting in the expression (\ref{gs}) for the normalized ground state.  
  
To construct the excited states of the CSM one  multiplies the ground state  
wave function by some symmetric polynomials  $\Psi (z) \sim J(z)\Psi_0 (z)$. The action of the   
kinetic term on this wave function generates the term (\ref{kinvdm1}) through the action of   
derivatives on $\ln \Psi_0 (z)$ and the term, which involves the derivatives of $\ln J(z)$,  
leading to the eigenvalue equation:  
\be  
\Big(H_0+\lambda H_1\Big) J(z) = \Big(E-E_0\Big) J(z)\, ,  
\label{hamJ}  
\ee  
where    
\be  
H_0 = \sum_i (z_i\d_i)^2 ,\;\;\;\;\;\;  
H_1 = \sum_{i\neq j}\frac{z_i+z_j}{z_i-z_j}(z_i\d_i-z_j\d_j) .  
						\label{h1}  
\ee  
Solutions to this differential equation        
$J (z) =  J^{1/\lambda}_{\prt{\kappa}}(z)$   are the symmetric Jack polynomials  
\cite{Macdonald} labeled by the quantum numbers  
$\prt{\kappa} = \prt{\kappa_1,\kappa_2,\ldots\kappa_N}$. These numbers are    
non-negative integers organized in the non-increasing sequence   
$\kappa_1\ge\kappa_2\ge\ldots\ge\kappa_N\ge 0$ are known as  parts  of the  
partition $\prt{\kappa}$. Since Jack polynomials are homogeneous functions,   
 under translations $z_j=e^{i\theta_j}\to u z_j=e^{i(\theta+\theta_j)}$  
they transform as   
\be  
J^{1/\lambda}_{\prt{\kappa}}(uz) = u^{|\kappa|}J^{1/\lambda}_{\prt{\kappa}}(z) .  
						\label{homo}  
\ee  
The total momentum of  an excited state is therefore   
\be  
P_\prt{\kappa} =P_0+  |\kappa|\, ,  
						\label{pkappa}  
\ee  
where $P_0 = \lambda N(N-1)/2$ is the total momentum of the ground state and   
$|\kappa| = \kappa_1+\kappa_2+\ldots +\kappa_N$.   
The energy  of the excitation  $\prt{\kappa}$ is given by  
\be  
E_\prt{\kappa} =  
E_0 +\sum_{j=1}^N \Big(\kappa_j^2 +2\lambda \kappa_j (N-j)\Big)\, .  
						\label{ekappa}  
\ee     
Each term in this sum  can be represented as an energy gain   
$\Big(\lambda (N-j) +\kappa_j\Big)^2-\Big (\lambda (N-j)\Big)^2$ of   
particle excited by momentum $\kappa_j$  from  its ground state momentum    $\lambda (N-j)$.

It is customary to represent partitions $\prt{\kappa}$ by  Young diagrams,  
a graph with $\kappa_j$ cells in the $j$-th row. A  dual partition is   
obtained by  
using the numbers of cells, $\tilde{\kappa}_a$, in the $a$-th column of the Young diagram.  

\begin{figure}     
\centerline{\psfig{figure=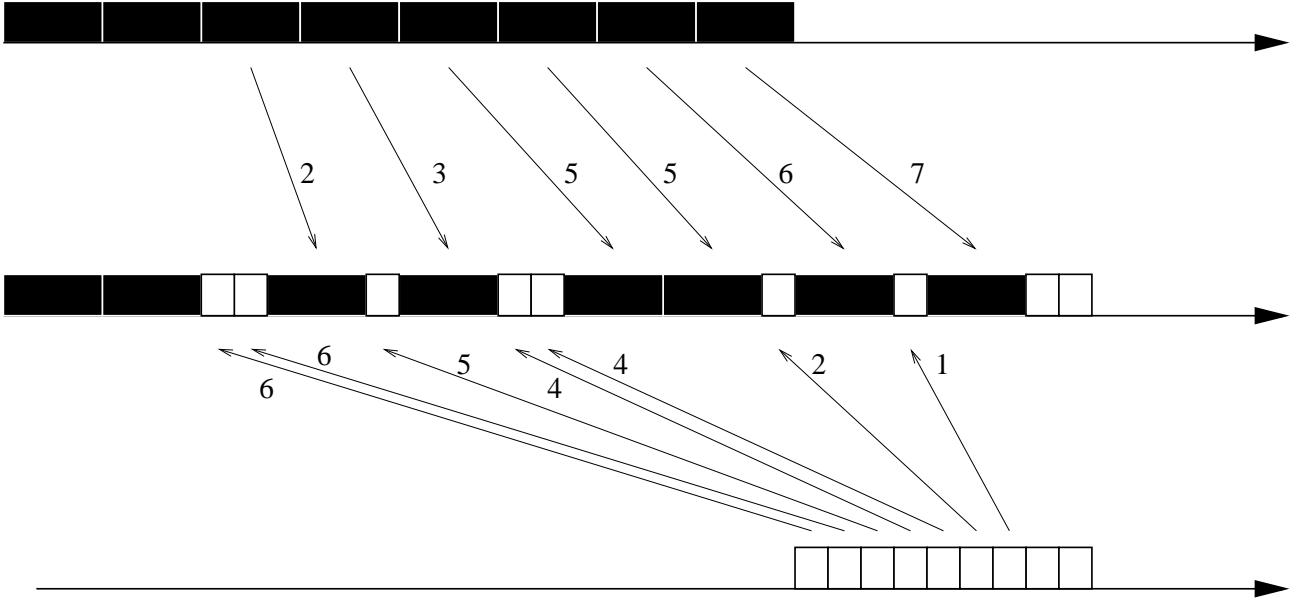,height=8cm} }    
\vspace{20pt}    
\caption{Another graphical representation of an excited state depicted in the previous figure. The excitations 
of particles and holes from their ground state position are related through the duality of the partitions. The 
number of particles is $N=8$ and the coupling constant $\lambda=3$.}     
\vspace{20pt}    
\label{bricks}  
\end{figure}

For example, the partition dual to $\prt{\kappa}=\prt{7,6,5,5,3,2}$   
( the one depicted in \fig{yd} ) is   
$\prt{\tilde{\kappa}}=\prt{6,6,5,4,4,2,1}$.   
It can be represented by transposing the  Young diagram for $\prt{\kappa}$.   
The dual partition  
can be given a nice interpretation in terms of hole excitations. Consider another graphical   
representation of the eigenstates of the Calogero-Sutherland model shown in \fig{bricks}.   
Each particle in the  ground state is represented by  black box of length $\lambda$,   
which corresponds to the   
momentum $\lambda (N-j)$ added to the ground state.  The excited state  
described by partition $\prt{\kappa}$ is represented by moving   
$j$-th particle   
($j=1$ corresponds to  the uppermost particle) by $\kappa_j$ units  up.   
Some holes of integer   
length 
will appear between the particles.   
Recording the shifts of the holes from their ground state positions,     
one finds that they are described by  the partition $\prt{\tkappa}$   
dual to $\prt{\kappa}$ as  
depicted in \fig{bricks}. In terms of the dual partition $\prt{\tkappa}$   
the energy (\ref{ekappa}) of the excitation reads  
\be  
E_\prt{\kappa} -E_0=   
\sum_a \Bigg( 2 \lambda\tkappa_a \Big(N+j/\lambda-(1+1/\lambda)/2\Big)-   
\lambda\tkappa_a^2\Bigg)\, ,  
						\label{etkappa}  
\ee   
while the total momentum is given by $P_\prt{\kappa}-P_0=|\tkappa|$, since the number of cells can  
be counted row-wise as well as column-wise. In the limit of large $N$   
the leading behavior of Eq.~(\ref{etkappa})  
is obtained by rescaling $\tkappa_a = Nk_a$ and reads  
$E_\prt{\kappa} -E_0\to N^2 \sum_a \lambda k_a(2-k_a)$.   
Here we must recall that we have chosen to work with the ground state   
with non--zero total momentum. That is equivalent to describing a system   
with zero--momentum ground state from the moving frame. Translation back to   
the frame where the ground state has zero momentum changes (only) expression   
for the excitation energy to  
\be  
E_\prt{\kappa} -E_0\to N^2 \sum_a \lambda k_a(1-k_a)\, .  
						\label{etk}  
\ee   
This is the expression we use to calculate the dynamical correlation function,  
which belongs therefore to the rest frame of the ground state.

The correspondence between dual partition and hole excitations is a non trivial fact and     
is crucial in our approach to the Calogero-Sutherland model.

\section{Matrix Elements}  
\label{app2}

Here we describe how various scalar products between symmetric functions   
can be represented in terms of partitions.   
Let us denote by   $j=1,\ldots,N$ the   
vertical coordinate and by $a=1,2,\ldots$ the horizontal    
coordinate of the cell in a Young diagram.   
The cell $(1,1)$ is the upper left corner of the     
partition. Then the symbols $\{\;\}^\lambda_\prt{\kappa}$ and     
$[\;\;]^\lambda_\prt{\kappa}$ are defined by the following product over     
all  cells in  the Young diagram for $\prt{\kappa}$:    
\begin{eqnarray}    
\label{curly}    
\{s\}^\lambda_\prt{\kappa} &\equiv& \prod_{j,a\in \prt{\kappa}}    
\Big(s-\lambda (j-1)+(a-1)\Big)     
\\    
\label{square}    
[s]^\lambda_\prt{\kappa} &\equiv& \prod_{j,a\in \prt{\kappa}}    
\Big(s+(a-1)/\lambda-(j-1)\Big) .    
\end{eqnarray}    
The Jack polynomials  are orthogonal with respect to the scalar product  
\be  
\int\limits_0^{2\pi}\! \frac{d^N\theta}{(2\pi)^N}   
\prod_{j<l} \left|e^{i\theta_l}-e^{i\theta_j}\right|^{2\lambda}  
J^{1/\lambda}_{\prt{\kappa}}(\bar{z})J^{1/\lambda}_{\prt{\kappa'}}(z)=  
\frac{j^\lambda_\prt{\kappa} [N]^\lambda_\prt{\kappa}}    
{[N+1/\lambda -1]^\lambda_\prt{\kappa}}\,   
\delta_{\prt{\kappa},\prt{\kappa'}}\, ,  
 						\label{scalar}  
\ee  
where the factor $j^\lambda_\prt{\kappa}$ may be written as the     
product over the cells of  the Young diagram:    
\be    
j^\lambda_\prt{\kappa}=\prod_{j,a\in \prt{\kappa}}     
\Big(\tkappa_a-j+(\kappa_j-a)/\lambda+1/\lambda\Big)    
\Big(\tkappa_a-j+(\kappa_j-a)/\lambda+1\Big) \, .    
\label{j}    
\ee    
Employing these expressions, one finds for the   
normalization constant $\N_\prt{\kappa}$ of excited states of the CSM  
(cf. Eq.~(\ref{es})):  
\be  
\N_\prt{\kappa} =\frac{[N+1/\lambda -1]^\lambda_\prt{\kappa}}  
{j^\lambda_\prt{\kappa} [N]^\lambda_\prt{\kappa}}\, .   
                                                 \label{nkappa}  
\ee

Jack polynomials form an orthogonal basis in the vector space of symmetric  
functions and can be used to expand any symmetric function. For example,   
the power  
sums, representing the Fourier coefficients of the density operator,    
Eq.~(\ref{fourcoeff}),   
are given by the following linear combination of Jack polynomials:  
\be  
\sum_j z_j^k = \frac{k}{\lambda}  
\sum_{|\kappa|=k}\frac{[0']^\lambda_\prt{\kappa}}{j^\lambda_\prt{\kappa}}  
J^{1/\lambda}_{\prt{\kappa}}(z)\, .  
						\label{powsumexp}  
\ee  
The prime in $ [0']^\lambda_\prt{\kappa}$ denotes that the cell $(1,1)$ is not included in the  
product in Eq.~(\ref{square}).   
More important for our discussion is the expansion of  
any power of the  characteristic polynomials:  
\be  
Z^s(0) =\prod_j (1-z_j)^s =    
\sum_\prt{\kappa}\frac{\{-s\}^\lambda_\prt{\kappa}}{\lambda^{|\kappa|}   
j^\lambda_\prt{\kappa}} J^{1/\lambda}_{\prt{\kappa}}(z)\, .  
						\label{chpolexp}  
\ee  
In terms of  normalized wave functions of the CSM, Eq.~(\ref{es}), the   
action of the $s$--power of  the characteristic polynomial operator   
on the ground state may be written as the following linear combination:  
\be  
Z^s(\theta)|0\> =  
\sum_\prt{\kappa} A_\prt{\kappa}(\lambda,N) \{-s\}^\lambda_\prt{\kappa}   
e^{-i|\kappa|\theta}|\prt{\kappa}\> \,  ,  
						\label{chpolexp1}  
\ee  
where  
\be  
A_\prt{\kappa}(\lambda,N) = \lambda^{-|\kappa|}  
\left(\frac {[N]^\lambda_\prt{\kappa}}  
{ j^\lambda_\prt{\kappa}[N+1/\lambda -1]^\lambda_\prt{\kappa}}\right)^{1/2}\ .  
						\label{coeff1}  
\ee

\section{The large $N$ limit}    
\label{app3}

\begin{figure}     
\centerline{\psfig{figure=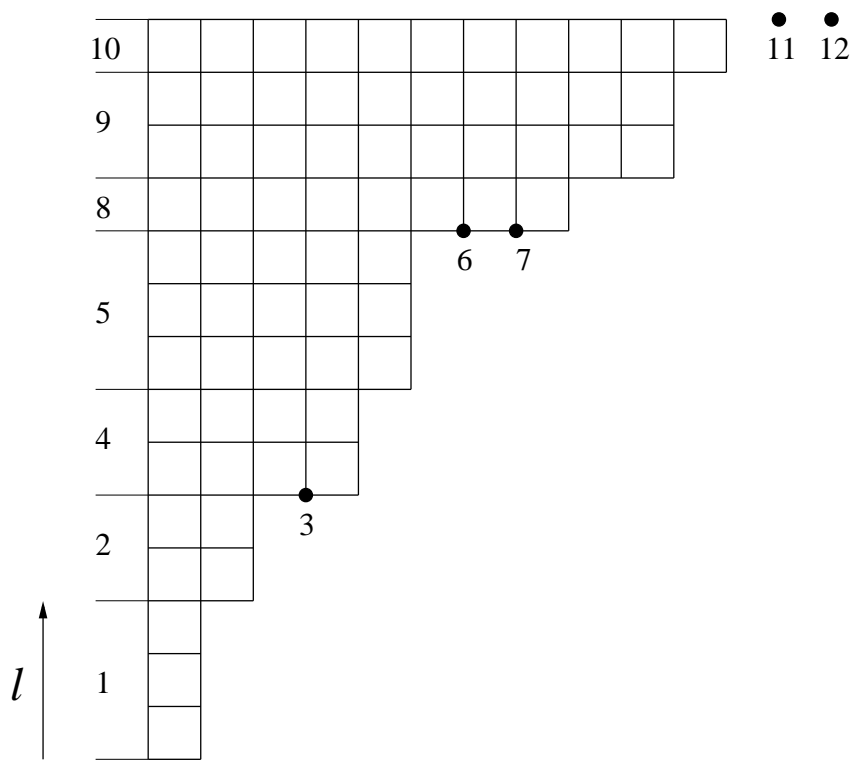,height=10cm} }    
\vspace{20pt}    
\caption{A Young diagram for $n=12$ divided in rectangles of fixed width $l$.     
Rectangles of zero height  $l=3,6,7,11,12$  are represented by dots.}     
\vspace{20pt}    
\label{div}     
\end{figure}

In this appendix we evaluate the thermodynamic limit $N\to\infty$ for the  
correlation function of the characteristic polynomials, Eq.~(\ref{corr}).   
To this end we need to consider the partition--dependent coefficient    
\be     
\frac{\lambda^{-2|\kappa|}}  {j^\lambda_\prt{\kappa}}\,     
\frac{ [N]^\lambda_\prt{\kappa}}    
{[N+1/\lambda -1]^\lambda_\prt{\kappa}}\,   
\left[\{-n\}^\lambda_\prt{\kappa}\right]^2     \, .    
                                                       \label{coeff2}    
\ee    
Let us start  with $\{-n\}^\lambda_\prt{\kappa}$     
given by the  product (\ref{curly}) over cells of the  Young diagram     
representing  $\prt{\kappa}$. It is clear     
from this definition  that $\{-n\}^\lambda_\prt{\kappa}=0$ if     
$\kappa_1>n$. One  can choose the heights   
$0\le \tkappa_a\le N\; ;\,\, a=1,\ldots,n$ of the first $n$     
columns as the coordinates, since only the partitions with at most $n$    
columns contribute.  In these coordinates one  can rewrite     
\be    
\{-n\}^\lambda_\prt{\kappa} =(-\lambda)^{|\kappa|}    
\prod_{a=1}^n \prod_{j=1}^{\tkappa_a}  \Big( (n-a+1)/\lambda+j-1\Big)     
=(-\lambda)^{|\kappa|}   
\prod_{a=1}^n \Big( (n-a+1)/\lambda\Big)_{\tkappa_a}\, .    
\label{mncurly}    
\ee    
Here we have used the definition of the Pochhammer's symbol:    
\be    
(z)_n \equiv z(z+1)(z+2)\ldots (z+n-1)=\frac{\Gamma(z+n)}{\Gamma(z)}\; ;  
\;\;\;\;\;    
(z)_0=1\, .    
\label{pochhammer}    
\ee    
As a result, one finds  
\be    
\left[\{-n\}^\lambda_\prt{\kappa}\right]^2=     
\lambda^{2|\kappa|} \prod_{a=1}^{n}          
\frac{\Gamma^2 \Big(\tkappa_a+(n-a+1)/\lambda\Big)}    
{\Gamma^2 \Big((n-a+1)/\lambda\Big)} \, .   
\label{mncurly2}    
\ee    
The factors $[N]^\lambda_\prt{\kappa}$ and     
$[N+1/\lambda -1]^\lambda_\prt{\kappa}$ are calculated in the same way by   
writing   the products, Eq.~(\ref{square}),  column-wise:    
\begin{eqnarray}    
\label{Nsquare1}    
 [N]^\lambda_\prt{\kappa} = \prod_{a=1}^n \prod_{j=1}^{\tkappa_a}     
\Big(N+(a-1)/\lambda-(j-1)\Big)   
&=&\prod_{a=1}^n     
\frac{\Gamma\Big(N+a/\lambda+1-1/\lambda\Big)}    
{\Gamma\Big(N-\tkappa_a+a/\lambda+1-1/\lambda\Big)}  \, ;  
\\    
\label{Nsquare2}    
 [N+1/\lambda-1]^\lambda_\prt{\kappa} =   
\prod_{a=1}^n \prod_{j=1}^{\tkappa_a}     
\Big(N+a/\lambda-j\Big)   
&=&\prod_{a=1}^n     
\frac{\Gamma\Big(N+a/\lambda\Big)}  
{\Gamma\Big(N-\tkappa_a+a/\lambda\Big)}\, .     
\end{eqnarray}      
Next we calculate the factor $j^\lambda_\prt{\kappa}$ defined in    
Eq.~(\ref{j}).     
We divide the Young diagram representing the partition $\prt{\kappa}$ into    
$n$ rectangles having the fixed width $b=1,2,\ldots,n$ and height     
$\tkappa_b-\tkappa_{b+1}$ as shown in \fig{div}. 
Note that     
when $ \tkappa_b=\tkappa_{b+1}$ the rectangle $b$     
has zero height. With these notations the expression    
(\ref{j})  can be written as a product over the rectangles:    
\bea    
j^\lambda_\prt{\kappa} &=&     
\prod_{b=1}^n \prod_{a=1}^b \prod_{j=\tkappa_{b+1}+1}^{\tkappa_b}     
\Big(\tkappa_a-j+(b-a+1)/\lambda\Big)    
\Big(\tkappa_a-j+1+(b-a)/\lambda\Big)     
\\    
\label{j1}    
&=&\prod_{b=1}^n \prod_{a=1}^b     
\frac{\Gamma\Big(\tkappa_a-\tkappa_{b+1}+(b-a+1)/\lambda\Big)\,    
\Gamma\Big(\tkappa_a-\tkappa_{b+1}+1+(b-a)/\lambda\Big)}    
{\Gamma\Big(\tkappa_a-\tkappa_b+(b-a+1)/\lambda\Big)\,    
\Gamma\Big(\tkappa_a-\tkappa_b+1+(b-a)/\lambda\Big)} \, .   
\eea    
Renaming the running index $b+1\to b$    
and changing the order of the products, the numerator of the last expression    
 may  be written as    
\bea    
&&\prod_{a=1}^n \Gamma\Big(\tkappa_a-\tkappa_{n+1}+(n+1-a)/\lambda\Big)\,    
\Gamma\Big(\tkappa_a-\tkappa_{n+1}+1+(n-a)/\lambda\Big)     
\\    
&\times&\prod_{b=a+1}^n      
\Gamma\Big(\tkappa_a-\tkappa_b-(a-b)/\lambda\Big)\,    
\Gamma\Big(\tkappa_a-\tkappa_b-(a-b)/\lambda+1-1/\lambda\Big)\; .    
\label{numj}    
\eea    
In the denominator of Eq.~(\ref{j1}) the order of the products is   
interchanged  and the terms with $a=b$ are factored out, the result is    
\be    
\Gamma^n(1/\lambda) \prod_{a=1}^n \prod_{b=a+1}^n     
\Gamma\Big(\tkappa_a-\tkappa_b-(a-b)/\lambda+1/\lambda\Big)\,    
\Gamma\Big(\tkappa_a-\tkappa_b-(a-b)/\lambda+1\Big) \; .    
\label{denj}    
\ee    
Finally, using the fact that $\tkappa_{n+1}=0$, one obtains     
\bea    
j^\lambda_\prt{\kappa} &=&     
\frac{1}{\Gamma^n(1/\lambda)} \prod_{a=1}^n    
\Gamma\Big(\tkappa_a+(n-a+1)/\lambda\Big)\,    
\Gamma\Big(\tkappa_a+1+(n-a)/\lambda\Big)     
\\    
\label{jfinal}    
&\times&\prod_{b=a+1}^n  \frac{    
\Gamma\Big(\tkappa_a-\tkappa_b-(a-b)/\lambda\Big)\,    
\Gamma\Big(\tkappa_a-\tkappa_b-(a-b)/\lambda+1-1/\lambda\Big)}{    
\Gamma\Big(\tkappa_a-\tkappa_b-(a-b)/\lambda+1/\lambda\Big)\,    
\Gamma\Big(\tkappa_a-\tkappa_b-(a-b)/\lambda+1\Big)}    
\; .    
\eea    
Combining the above  expressions together one finally obtains for the   
coefficient Eq.~(\ref{coeff2}):   
\bea      
\prod_{a=1}^n &&   
\frac{\Gamma(1/\lambda)}{\Gamma^2\Big((n-a+1)/\lambda\Big) }    
\\    
\nn    
\times \prod_{a=1}^n &&     
\frac{\Gamma \Big(\tkappa_a+(n-a)/\lambda+1/\lambda\Big)}    
{\Gamma\Big(\tkappa_a+(n-a)/\lambda+1\Big) }\,      
\frac{\Gamma\Big(N-\tkappa_a+a/\lambda\Big)}    
{\Gamma\Big(N-\tkappa_a+a/\lambda+1-1/\lambda\Big)}\,     
\frac{\Gamma\Big(N+a/\lambda+1-1/\lambda\Big)}{\Gamma\Big(N+a/\lambda\Big)}    
\\    
\times\prod_{b=a+1}^n   &&     
\frac{\Gamma\Big(\tkappa_a-\tkappa_b-(a-b)/\lambda+1/\lambda\Big)\,    
\Gamma\Big(\tkappa_a-\tkappa_b-(a-b)/\lambda+1\Big)}{    
\Gamma\Big(\tkappa_a-\tkappa_b-(a-b)/\lambda\Big)\,    
\Gamma\Big(\tkappa_a-\tkappa_b-(a-b)/\lambda+1-1/\lambda\Big)} \; .    
\label{coeffinal}    
\eea     
The thermodynamic limit, $N\to\infty$,    
is obtained with the help of the following asymptotic relation:    
\be    
\lim_{|z|\to\infty}\frac{\Gamma(z+a)}{\Gamma (z)}= z^a \; .    
\label{assrel}    
\ee    
Rescaling the column heights as $\tkappa_a=N k_a$,  
one obtains for the thermodynamic limit of the expression (\ref{coeffinal})    
\be     
\frac{N^{\frac{n^2}{\lambda}}}{N^n}\,   
\prod_{a=1}^n \frac{\Gamma(1/\lambda)}    
{\Gamma^2\Big((n-a+1)/\lambda\Big) }    
\prod_{1\le a<b\le n} (k_a-k_b)^{2/\lambda}    
\prod_{a=1}^n k_a^{1/\lambda-1} (1-k_a)^{1/\lambda-1}\; .    
\label{coefftherm}    
\ee    
We relax the requirement $k_1\ge k_2\ge\ldots\ge k_n$     
allowing all the variables to run independently between $0$ and $1$.   
To do so  we have to  divide by $n!$ and  take the     
absolute value of the Vandermonde determinant:    
\be     
\prod_{1\le a<b\le n} (k_a-k_b)^{2/\lambda}\to      
|\Delta_n(k)|^{2/\lambda}   \, .  
\label{vandermondek}    
\ee    
Finally, changing the sum over partitions    
to the integrals over $k_a$ and       
absorbing the factor $N^{-n}$ into the integration measure,   
one recovers Eqs.~(\ref{corrtherm}) and (\ref{c}).

\end{document}